\documentclass[10pt,a4paper,fleqn]{article}

\usepackage[T1]{fontenc}
\usepackage[latin1]{inputenc}
\usepackage{wasysym}
\usepackage{graphicx}
\usepackage{natbib}
\usepackage{txfonts}

\newcommand{\eq}[1]{\begin{equation}  #1 \end{equation}}

\newcommand{\eqa}[1]{\begin{eqnarray}   #1 \end{eqnarray}}

\newcommand{\br}[1]{\left( #1 \right)}
\newcommand{\bc}[1]{\left\{ #1 \right\}}
\newcommand{\bb}[1]{\left[ #1 \right]}
\newcommand{\ba}[1]{\left\langle #1 \right\rangle}

\newcommand{\nn}{\nonumber}

\newcommand{\dd}{{\rm d}}

\begin{document}
\begin{center}
{\huge \textbf{Controlling intrinsic alignments in\\[.5ex] weak lensing statistics}}\\[2ex]
{\Large \textbf{The nulling and boosting techniques}}\\[2ex]
{\large B. Joachimi and P. Schneider}
\end{center}
\vspace*{0.3cm}
Argelander-Institut f\"ur Astronomie (AIfA), Universit\"at Bonn, Auf dem H\"ugel 71, 53121 Bonn, Germany\\
\textit{email}: {\tt joachimi@astro.uni-bonn.de}\\[2ex]

\begin{center}
\begin{minipage}[c]{.75\textwidth}
The intrinsic alignment of galaxies constitutes the major astrophysical source of systematic errors in surveys of weak gravitational lensing by the large-scale structure. We discuss the principles, summarise the implementation, and highlight the performance of two model-independent methods that control intrinsic alignment signals in weak lensing data: the nulling technique which eliminates intrinsic alignments to ensure unbiased constraints on cosmology, and the boosting technique which extracts intrinsic alignments and hence allows one to further study this contribution. Making only use of the characteristic dependence on redshift of the signals, both approaches are robust, but reduce the statistical power due to the similar redshift scaling of intrinsic alignment and lensing signals.
\end{minipage}
\end{center}
\vspace*{0.5cm}

\section{Cosmic shear and intrinsic alignments}

The weak gravitational lensing of distant galaxies by the large-scale structure, or cosmic shear in short, is one of the most powerful cosmological probes of dark matter, dark energy, deviations from general relativity, or the initial conditions of structure formation \citep[e.g.][]{peacock06,albrecht06}. Recent observational results \citep[e.g.][]{benjamin07,schrabback09} demonstrate the increasing maturity and competitiveness of cosmic shear, while ambitious planned survey such as LSST\footnote{\texttt{http://www.lsst.org}} and Euclid\footnote{\texttt{http://sci.esa.int/science-e/www/area/index.cfm?fareaid=102}} feature cosmic shear as one of their primary probes.

The gradually tightening statistical constraints on cosmology by upcoming cosmic shear surveys entail more and more stringent requirements on the control of systematic errors. While instrumental effects, which can e.g. affect the measurement of galaxy shapes, are eliminated via specific instrument designs, survey strategies, or steps in the reduction pipeline \citep[see e.g.][]{bridle09}, astrophysical sources of systematics have to be dealt with at the data level. The major astrophysical systematic in cosmic shear surveys is generated by the intrinsic alignment of galaxy shapes, which we will consider in detail in the following.

At the two-point level cosmic shear measures, such as the correlation functions or the power spectrum, are based on the correlator of the (complex) measured ellipticity $\epsilon$ of pairs of galaxies \citep[for an overview on weak lensing theory see][]{bartelmann01}. In the limit of very weak lensing this ellipticity is given by the sum of the intrinsic ellipticity of a galaxy $\epsilon^{\rm s}$ and the gravitational shear $\gamma$. Consequently, a two-point correlator of galaxy samples $i$ and $j$ has contributions by four terms,
\eqa{
\label{eq:epscorrelators}
\hspace*{-0.8cm} \ba{\epsilon_i \epsilon_j^*} &=& \underbrace{\ba{\gamma_i \gamma_j^*}}+\underbrace{\ba{\epsilon_i^{\rm s} \epsilon_j^{{\rm s}*}}}+\underbrace{\ba{\gamma_i \epsilon_j^{{\rm s}*}}+\ba{\epsilon_i^{\rm s} \gamma_j^*}}\;,\\ \nn
\hspace*{-0.8cm} && \hspace*{0.2cm} {\rm GG} \hspace*{0.9cm} {\rm II} \hspace*{1.6cm} {\rm GI}
}
where the first term on the right-hand side is the desired cosmic shear signal (GG henceforth). The second term denotes correlations between the intrinsic ellipticities of galaxies (II henceforth), and the third term comprises correlations between the intrinsic ellipticity of one galaxy and the gravitational shear acting on another (GI henceforth). In standard cosmic shear analysis it is assumed that the intrinsic shape of a galaxy is not correlated with either the intrinsic shape or the gravitational shear acting on another galaxy. However, galaxies do indeed show intrinsic alignments, which thus induces systematic errors.

A straightforward solution to this issue would be the modelling of the II and GI signals, possibly including a set of nuisance parameters in the cosmic shear likelihood analysis, but this approach is only safe if the functional form of the systematic is well known \citep[e.g.][]{kitching08}. Unfortunately, the understanding of intrinsic alignments is currently still at a crude level because it involves the intricacies of galaxy formation and evolution in the local large-scale environment, including the complications by baryonic physics. Analytic modelling of intrinsic alignments is only applicable on relatively large scales (see \citealp{schneiderm09} for recent developments), and simulations are either limited by simplistic models of placing galaxies into dark matter haloes \citep[e.g.][]{heymans06,semboloni08}, or by the small volume if baryons are included \citep[e.g.][]{bett10,hahn10}.

Observations of intrinsic alignments, using shallow surveys in which cosmic shear is subdominant or cross-correlations between galaxy number densities and shapes \citep[e.g.][]{mandelbaum06,hirata07,mandelbaum09,joachimi10c}, have revealed that this systematic depends on galaxy colours (and hence type) as well as galaxy luminosity (and hence mass). However, constraints on specific intrinsic alignment models are still weak and inherit systematic uncertainties from e.g. galaxy luminosity functions or $k$-corrections. Furthermore, these observations are only possible for galaxy samples which are either luminous or at low redshift, and therefore can have spectroscopic or high-quality photometric redshift measurements, so that a substantial amount of extrapolation is needed to predict intrinsic alignment signals from the typically faint and high-redshift galaxies in cosmic shear surveys.

Hence, model-independent methods that remove the contamination by intrinsic alignments from cosmic shear surveys are desirable. Besides, it would be convenient to be able to extract the intrinsic alignment signal from the cosmic shear survey itself in order to at least provide a cross-check for the extrapolation of intrinsic alignment models from brighter and low-redshift samples. These two goals are met by the nulling and boosting techniques which will be presented in the following, based on the work by \citet[Paper I, II, III hereafter]{joachimi08b,joachimi09,joachimi10b}.

\section{The principle of nulling and boosting}
\label{sec:principle}

As cosmic shear two-point statistics are based on the correlator (\ref{eq:epscorrelators}), all of them are affected by additive contributions from II and GI correlations, so that for the observed angular power spectrum\footnote{Since the methods which are going to be discussed do not depend on angular scales, the following considerations hold for any cosmic shear two-point statistic. We use power spectra throughout because they are the easiest to handle computationally.} one obtains
\eq{
\label{eq:signals}
\hspace*{-0.8cm} C_{\rm obs}^{(ij)}(\ell) = C_{\rm GG}^{(ij)}(\ell) + C_{\rm II}^{(ij)}(\ell) + C_{\rm GI}^{(ij)}(\ell)\;,
}
where the superscripts in parentheses indicate that we consider tomographic cosmic shear data with additional photometric redshift information \citep[see e.g.][]{schrabback09}, correlating photometric redshift bin $i$ with bin $j$. All power spectra on the right-hand side of (\ref{eq:signals}) are given in terms of their underlying three-dimensional power spectra by the following Limber equations \citep[e.g.][]{hirata04,bridle07,joachimi10},
\eqa{
\label{eq:limber}
\hspace*{-0.8cm} C_{\rm GG}^{(ij)}(\ell) &=& \int^{\chi_{\rm hor}}_0 \dd \chi\; \frac{q^{(i)}(\chi)\, q^{(j)}(\chi)}{\chi^2}\; P_{\delta} \br{\frac{\ell}{\chi},\chi}\;;\\ 
\label{eq:limberII}
\hspace*{-0.8cm} C_{\rm II}^{(ij)}(\ell) &=& \int^{\chi_{\rm hor}}_0 \dd \chi\; \frac{p^{(i)}(\chi)~p^{(j)}(\chi)}{\chi^2}\; P_{\rm II} \br{\frac{\ell}{\chi},\chi}\;;\\ 
\label{eq:limberGI}
\hspace*{-0.8cm} C_{\rm GI}^{(ij)}(\ell) &=& \int^{\chi_{\rm hor}}_0 \dd \chi\; \frac{p^{(i)}(\chi)~q^{(j)}(\chi) + q^{(i)}(\chi)~p^{(j)}(\chi)}{\chi^2}\; P_{\delta {\rm I}} \br{\frac{\ell}{\chi},\chi}\;.
}
Here, $P_{\delta}$ denotes the matter power spectrum, $P_{\rm II}$ the intrinsic ellipticity power spectrum, and $P_{\delta {\rm I}}$ the cross-power spectrum between the matter field and the distribution of intrinsic galaxy shapes. The latter two have to be specified by an intrinsic alignment model, but are generally unknown. Note that we assume a spatially flat universe throughout.

The integrals in (\ref{eq:limber}) to (\ref{eq:limberGI}) run over the comoving distance $\chi$ out to the comoving distance horizon $\chi_{\rm hor}$. The kernels are composed of the probability distribution of comoving distances $p^{(i)}(\chi)$ in photometric redshift bin $i$ and the lensing weight
\eq{
\label{eq:lenseff}
\hspace*{-0.8cm} q^{(i)}(\chi) = \frac{3H_0^2 \Omega_{\rm m}}{2 c^2}\; \chi\; \bb{1+z(\chi)} \int_\chi^{\chi_{\rm hor}} \dd \chi'\, p^{(i)}(\chi')\, \br{1 - \frac{\chi}{\chi'}}\;,
}
where the integral corresponds to the typical scaling of shear signals with the average over the source galaxy distribution $i$ of the ratio of distance between lens and source over the distance between observer and source. The distributions $p^{(i)}(\chi)$ are usually chosen to be narrow in cosmic shear tomography, whereas the lensing weight (\ref{eq:lenseff}) is broad. These two quantities determine the scaling of the power spectra (\ref{eq:limber}) to (\ref{eq:limberGI}) with redshift, which holds in particular for the II and GI signals, if $p^{(i)}(\chi)$ is compact enough that the evolution of the three-dimensional intrinsic alignment power spectra over the integration range is negligible.

\begin{figure}[t]
\begin{minipage}[c]{.6\textwidth}
\centering
\includegraphics[scale=.4,angle=270]{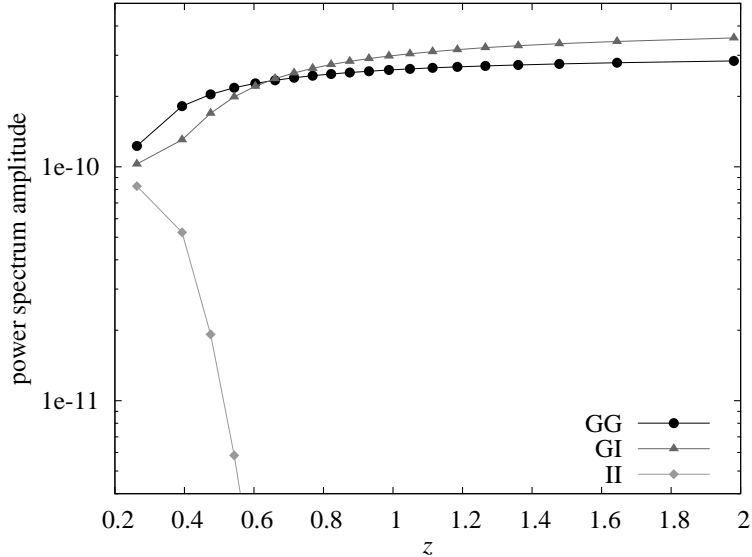}
\end{minipage}%
\begin{minipage}[c]{.4\textwidth}
\caption{Redshift dependence of the GG, GI, and II signals. Shown is the amplitude of the power spectra $C^{(i=1,\,j)}(\ell)$ at $\ell \approx 200$ as a function of the median redshift of galaxy sample $j$. The power spectra are computed for a $\Lambda$CDM cosmology and a Euclid-like redshift distribution divided into 20 bins containing an equal number of galaxies, and using a photometric redshift dispersion of $0.05(1+z)$. The GI and II terms are obtained from the non-linear version of the linear alignment model. The GG signal is shown as black circles, the GI signal as dark grey triangles, and the II signal as light grey diamonds.}
\label{fig:zdep}
\end{minipage}
\end{figure}

As a first step, we illustrate the redshift dependence of the GG, GI, and II signals. To compute the matter power spectrum, we assume a flat $\Lambda$CDM universe with parameters $\Omega_{\rm m}=0.25$, $\sigma_8=0.8$, $n_{\rm s}=1$, $h=0.7$, and $\Omega_{\rm b}=0.05$, using the transfer function by \citet{eisenstein98} and the non-linear corrections provided by \citet{smith03}. The two intrinsic alignment power spectra are determined via the non-linear version of the linear alignment model \citep{catelan01,hirata04,bridle07}, with the normalisation given in \citet{bridle07}. We emphasise that an intrinsic alignment model is only introduced for illustrational purposes and to assess the performance of the methods we propose, while the implementation of these techniques does not depend on any knowledge about intrinsic alignments other than the well-known form of (\ref{eq:limberII}) and (\ref{eq:limberGI})\footnote{Recent observational results \citep{mandelbaum09,joachimi10c} suggest that the model we use over-predicts the intrinsic alignment signal of typical galaxies in deep cosmic shear surveys. In addition, the implemented model has been shown in the meantime to contain a mathematical error (Hirata 2010, in prep.; \citealp{joachimi10c}). Both issues do not have a significant impact on the interpretation of our results.}. Here and in the following we will assume a Euclid-like survey covering $20000\,{\rm deg}^2$ with a median redshift of $0.9$ and a photometric redshift scatter of $\sigma_{\rm ph}(1+z)$ with $\sigma_{\rm ph}=0.05$ \citep[for details see][]{laureijs09}.

In Fig.$\,$\ref{fig:zdep} we have plotted the angular power spectra $C^{(ij)}(\ell)$ for $i=1$ and $\ell \approx 200$ fixed as a function of the median redshift of the photometric redshift bin $j$. We have split the survey into $N_z=20$ photometric redshift bins such that each bin contains the same number of galaxies. It is evident that the II signal has a distinctive dependence on redshift, quickly decreasing in amplitude if the distance between bins $i$ and $j$ increases. This behaviour is related to the narrow kernel in (\ref{eq:limberII}) and originates from the fact that galaxies have to be physically close in order to be aligned by the same matter structures. Thus the II signal is relatively straightforward to remove or isolate, also in a model-independent way \citep{king02,king03,heymans03,takada04b}. In contrast, the GI and GG signals display a very similar scaling with redshift as both have lensing contributions. In the following we will therefore concentrate on the GI signal, discussing the treatment of II correlations mostly in a qualitative manner.

To construct methods that make use of the characteristic dependence on redshift of the GI and GG signal displayed in Fig.$\,$\ref{fig:zdep}, we first consider the limit of very narrow redshift bins, i.e. $p^{(i)}(\chi) \approx \delta_{\rm D}(\chi -\chi_i)$, where $\delta_{\rm D}$ is the Dirac delta-distribution, and where $\chi_i$ is a comoving distance on which $p^{(i)}(\chi)$ is centred. The lensing weight (\ref{eq:lenseff}) can then be written as
\eq{
\label{eq:lensefftransition}
\hspace*{-0.8cm} q^{(i)}(\chi) \rightarrow q(\chi_i,\chi) \equiv \frac{3H_0^2 \Omega_{\rm m}}{2 c^2}\; \chi\; \bb{1+z(\chi)}\; \left\{ 
\begin{array}{ll}
1 - \frac{\chi}{\chi_i}    &~~\mbox{if}~~ \chi < \chi_i\\
0                          &~~\mbox{else}\;.  
\end{array} \right.
}
With this equation at hand, the angular GG and GI power spectra turn into
\eqa{
\label{eq:GGapprox}
\hspace*{-0.8cm} C_{\rm GG}(\chi_i,\chi_j,\ell) &=&  \int^{{\rm min}(\chi_i,\chi_j)}_0 \dd \chi\; \frac{q(\chi_i,\chi)\; q(\chi_j,\chi)}{\chi^2}\; P_{\delta} \br{\frac{\ell}{\chi},\chi}\;;\\ 
\label{eq:GIapprox}
\hspace*{-0.8cm} C_{\rm GI}(\chi_i,\chi_j,\ell) &=& q(\chi_j,\chi_i)\; \chi_i^{-2}\; P_{\delta {\rm I}} \br{\frac{\ell}{\chi_i},\chi_i} + q(\chi_i,\chi_j)\; \chi_j^{-2}\; P_{\delta {\rm I}} \br{\frac{\ell}{\chi_j},\chi_j}\;,
}
where we have made the dependence on $\chi_i$ and $\chi_j$ explicit in the arguments. Note that the first term in (\ref{eq:GIapprox}) only contributes if $\chi_i < \chi_j$, and vice versa for the second term. Note furthermore that in this approximation the dependence of $C_{\rm GI}$ on the background bin (e.g. bin $j$ in the first term) is only through the lensing weight $q$.

We now define transformed power spectra as simple linear combination of the original angular power spectra, summing over the index of the background photometric redshift bin,
\eq{
\label{eq:nullingtrafo}
\hspace*{-0.8cm} \Pi^{(i)}_{[m]}(\ell) = \sum_{j=j_{\rm min}}^{N_z} {T_{[m]}^{(i)}}_j ~C_{\rm obs}^{(ij)}(\ell)\;,
}
for every foreground bin $i$, where the sum starts at the bin index $j_{\rm min}$. We are going to determine the weights $T_{[m]}^{(i)}$ such that either the GI signal is eliminated (\lq nulled\rq), or the GG signal is suppressed (and GI correspondingly \lq boosted\rq).

Retaining the approximation of narrow redshift bins, one obtains by inserting (\ref{eq:GIapprox}) into (\ref{eq:nullingtrafo}) that the GI signal is removed for $j_{\rm min} \geq i$ if the weights fulfil the condition
\eq{
\label{eq:nullingcondition}
\hspace*{-0.8cm} \sum_{j=j_{\rm min}}^{N_z} {T_{[m]}^{(i)}}_j\; \br{1 - \frac{\chi_i}{\chi_j}} = 0\;.
}
Since auto-correlations with $i=j$ are likely to be contaminated by an II term, we discard them completely and set $j_{\rm min} = i+1$. Then $N_z-i$ weights enter a linear combination (\ref{eq:nullingtrafo}) which only has to fulfil the condition (\ref{eq:nullingcondition}), so that several such linear combinations can be constructed, indexed by the subscript $[m]$. 

To obtain the central condition for boosting, we insert (\ref{eq:GGapprox}) into (\ref{eq:nullingtrafo}) instead, finding that the GG signal is eliminated if
\eqa{
\label{eq:boostingcondition}
\hspace*{-0.8cm} && \frac{3H_0^2 \Omega_{\rm m}}{2 c^2} \int^{{\rm min}(\chi_i,\chi_j)}_0 \dd \chi\; \frac{q(\chi_i,\chi)\; G^{(i)}(\chi)}{\chi}\; \bb{1+z(\chi)}\; P_{\delta} \br{\frac{\ell}{\chi},\chi}\\ \nn
\hspace*{-0.8cm} &\approx& \frac{9H_0^4 \Omega_{\rm m}^2}{4 c^4} \int_0^{\chi_i} \dd \chi \br{1 - \frac{\chi}{\chi_i}} G^{(i)}(\chi) = 0
}
is satisfied. Here, we defined the function
\eq{
\label{eq:boostingG}
\hspace*{-0.8cm} G^{(i)}(\chi_k) = \sum_{j=k}^{N_z} {T_{[m]}^{(i)}}_j\; \br{1 - \frac{\chi_k}{\chi_j}}\;
}
in a discretised form which can be used in (\ref{eq:boostingcondition}) if the integral is approximated by its Riemannian sum. To arrive at the second equality in (\ref{eq:boostingcondition}), we inserted (\ref{eq:lensefftransition}) and assumed that the redshift evolution of the matter power spectrum is given by $P_\delta (k,z) \propto (1+z)^{-2}$ which holds in the matter-dominated era. Since the matter power spectrum is already known to fairly high precision, one could use the first term in (\ref{eq:boostingcondition}) directly, but we choose to work with the second term to ensure a maximum of model independence. Since the weights $T_{[m]}^{(i)}$ define $G^{(i)}$ for a set of arguments $\chi_k$, there is no additional freedom (except for the arbitrary normalisation of the weights) to construct further linear combinations, contrary to the case of nulling. Hence we will use a single set of weights to extract the GI signal in redshift bin $i$. 

Note that the weights depend on comoving distances via (\ref{eq:nullingcondition}) and (\ref{eq:boostingG}), while the observed signals $C_{\rm obs}^{(ij)}(\ell)$ are binned in terms of redshift. Therefore the distance-redshift relation and consequently cosmological parameters like $\Omega_{\rm m}$, or the dark energy equation-of-state parameters, enter the weights $T_{[m]}^{(i)}$, but this dependence is weak, and even grossly incorrect a-priori assumptions for the values of these parameters are readily corrected for in an iterative scheme (see Paper II).

\section{Nulling intrinsic alignments}

\begin{figure}[t]
\begin{minipage}[c]{.6\textwidth}
\centering
\includegraphics[scale=.4,angle=270]{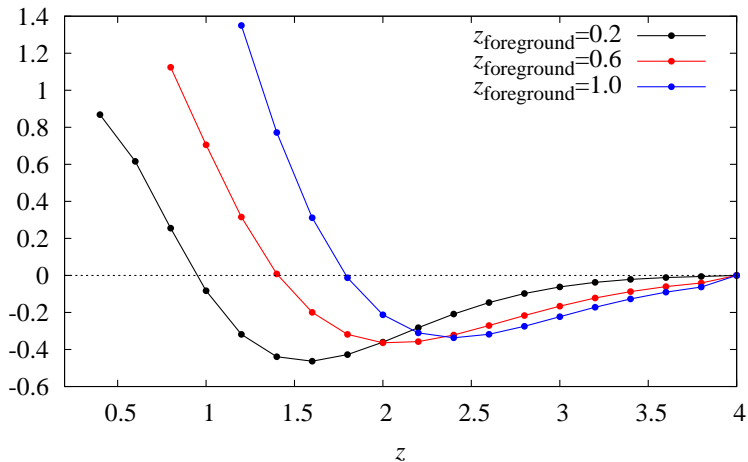}
\end{minipage}%
\begin{minipage}[c]{.4\textwidth}
\caption{Example set of nulling weights which additionally have been optimised to retain a maximum of cosmological information. The weights have been obtained for an equidistant redshift binning with $N_z=20$ between $z=0.2$ and $z=4$. Weights are shown for the foreground bins $i=1$ (in black), $i=3$ (in red), and $i=5$ (in blue), with the lower redshift boundary of these foreground bins given in the legend.}
\label{fig:nullingweights}
\end{minipage}
\end{figure}

The additional freedom in the determination of the nulling weights (\ref{eq:nullingcondition}) can be exploited to maximise the information on cosmology contained in the transformed measures (\ref{eq:nullingtrafo}). As investigated in Paper I, this can be achieved by calculating the weights such that a functional of the Fisher matrix like the trace or determinant is maximised. The choice of functional or Fisher matrix element does not have a strong effect on the resulting form of the weights, see e.g. \citet{shi10}. Higher-order modes of the transformed power spectra $\Pi^{(i)}_{[m]}(\ell)$ with $m \geq 1$ are then obtained by requiring that the corresponding weights be orthogonal with respect to the weights of all lower-order modes. We showed in Paper I that nearly all cosmological information is already contained in the first order $m=1$. In Fig.$\,$\ref{fig:nullingweights} an exemplary set of weights $T_{[1]}^{(i)}$ for three different bins $i$ has been plotted. Due to the nulling condition (\ref{eq:nullingcondition}), these weights necessarily have a zero crossing. They are largest if foreground bin and background bin are close in redshift where the GI signal is small, and have the smallest absolute values at high redshift were the GI term is large relative to the lensing contribution.

Alternatively, one can drop the condition on maximising cosmological information and simply compute sets of weights that are mutually orthogonal. For $N_z-i$ power spectra entering (\ref{eq:nullingtrafo}) for a given $i$ there are just $N_z-i$ such sets of weights where one, however, cannot fulfil the condition (\ref{eq:nullingcondition}) anymore, so that the corresponding transformed power spectrum is discarded in the further analysis. Imposing in addition unit normalisation on these sets of weights, this approach allows for the convenient interpretation that nulling is equivalent to a rotation of the cosmic shear data vector, with a subsequent truncation of certain contaminated elements in the rotated data vector. This ansatz was investigated in Paper II and is computationally preferable because it does not require Fisher matrices to determine the nulling weights, but only (\ref{eq:nullingcondition}) and a standard orthogonalisation procedure. Note that in this case no particular order is assigned to the modes $m$.

\begin{figure}[t!!!]
\begin{minipage}[c]{.5\textwidth}
\centering
\includegraphics[scale=.35,angle=270]{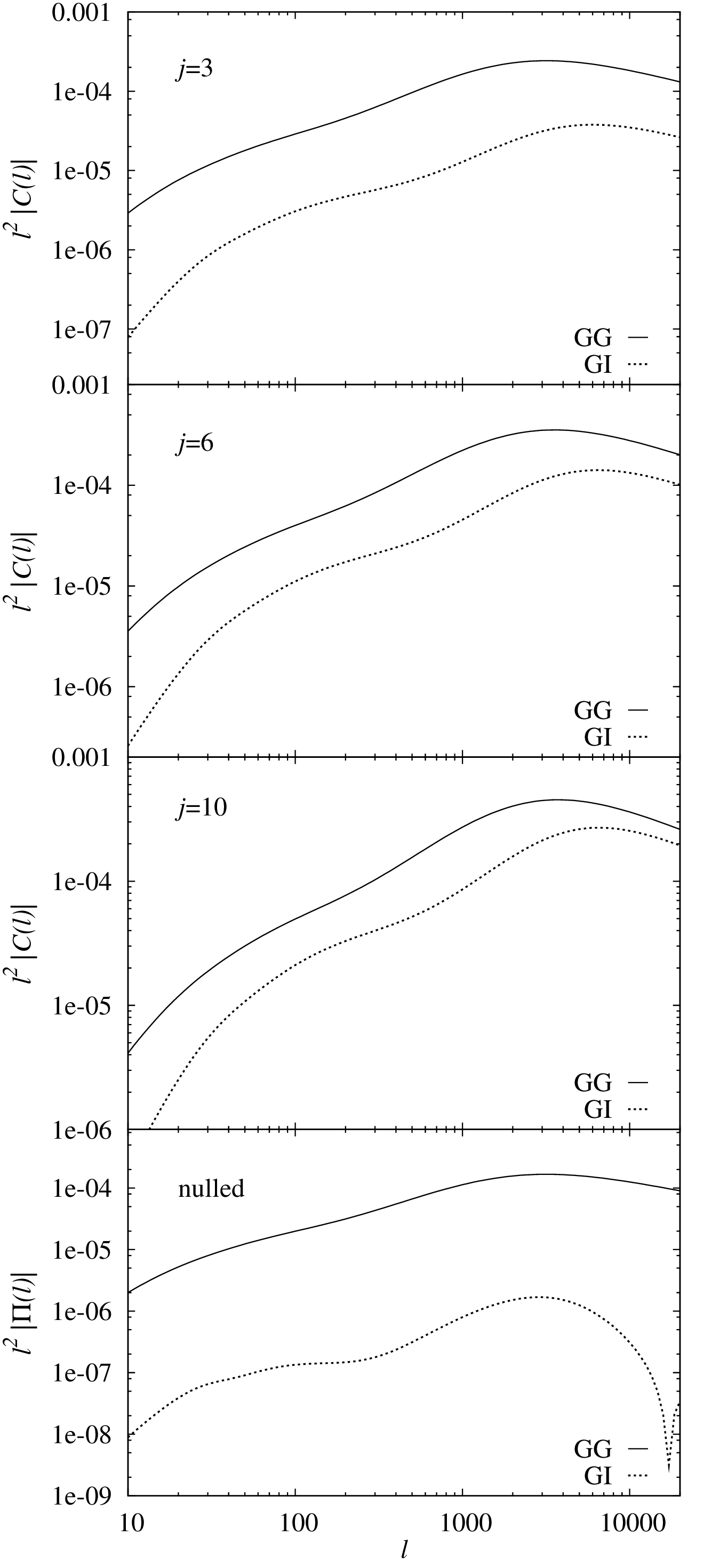}
\end{minipage}%
\begin{minipage}[c]{.5\textwidth}
\includegraphics[scale=.35,angle=270]{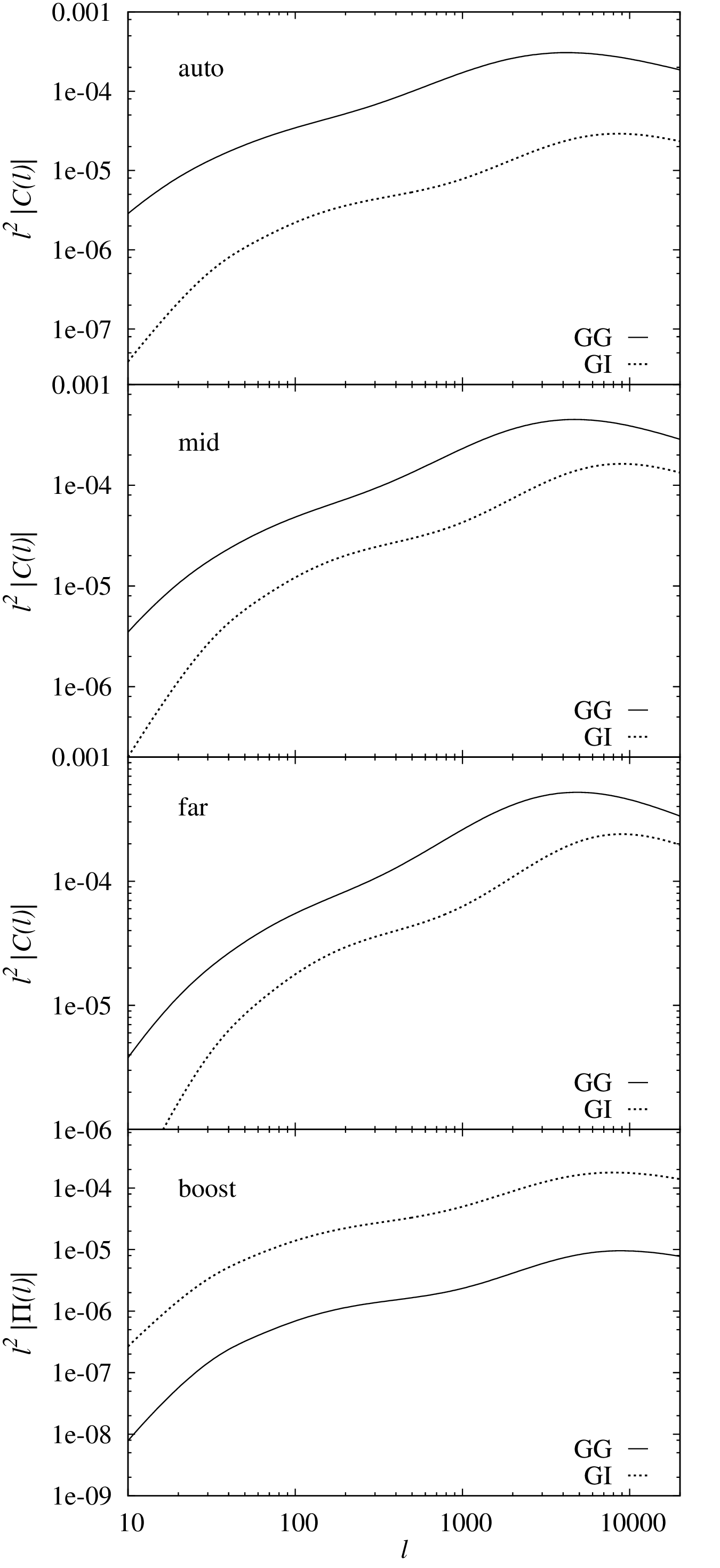}
\end{minipage}
\caption{\textit{Left panel}: Effect of nulling on GG and GI power spectra for a Euclid-like survey divided into $N_z=10$ bins, each containing the same number of galaxies, and with $\sigma_{\rm ph}=0.03$. Shown are the original GG (solid lines) and GI (dotted lines) power spectra for foreground bin $i=3$ and background bin $j=3$ (top panel), $j=6$ (upper centre panel), and $j=10=N_z$ (lower centre panel). The bottom panel displays the nulled power spectra for $i=3$ and $m=1$. \textit{Right panel}: Effect of boosting on GG and GI power spectra, also for $\sigma_{\rm ph}=0.03$, and obtained for the default binning used. The coding of the curves is the same as above. The upper three panels show power spectra for a foreground bin at $z_i=0.76$ and different background bins $j$, i.e. auto-correlations ($j=i$, \lq auto\rq), cross-correlations with a bin at intermediate redshift ($j=(i+N_z)/2$, \lq mid\rq), and cross-correlations with the most distant bin ($j=N_z$, \lq far\rq). In the bottom panel the boosted GG and GI signals are plotted. Note that absolute values of the power spectra are shown throughout.}
\label{fig:ps}
\end{figure}

In Fig.$\,$\ref{fig:ps}, left panel, the effect of the nulling transformation on power spectra is demonstrated, assuming a Euclid-like survey as detailed in the foregoing section, but for $N_z=10$ and a lower photometric redshift scatter $\sigma_{\rm ph}=0.03$\footnote{Note that the analysis in this section is based on a slightly different cosmology with $\sigma_8=0.9$, using the non-linear corrections to the matter power spectrum by \citet{PeacockDodds}. These differences should have a very minor influence on the compatibility of our results.}. While for the original power spectra $C^{(ij)}(\ell)$ the absolute value of the GI signal attains about $10\,\%$ (for auto-correlations, $i=j$) to $50\,\%$ (for a high redshift of the background bin, $j=N_z$), the GI term is suppressed by roughly two orders of magnitude with respect to GG in the corresponding nulled power spectrum. Depending on the mode $m$, the residual GI signal can feature one (as in the depicted case) or more zero crossings and hence acts more like a noise contribution (see Paper II).

\begin{figure}[t]
\begin{minipage}[c]{.72\textwidth}
\centering
\includegraphics[scale=.6]{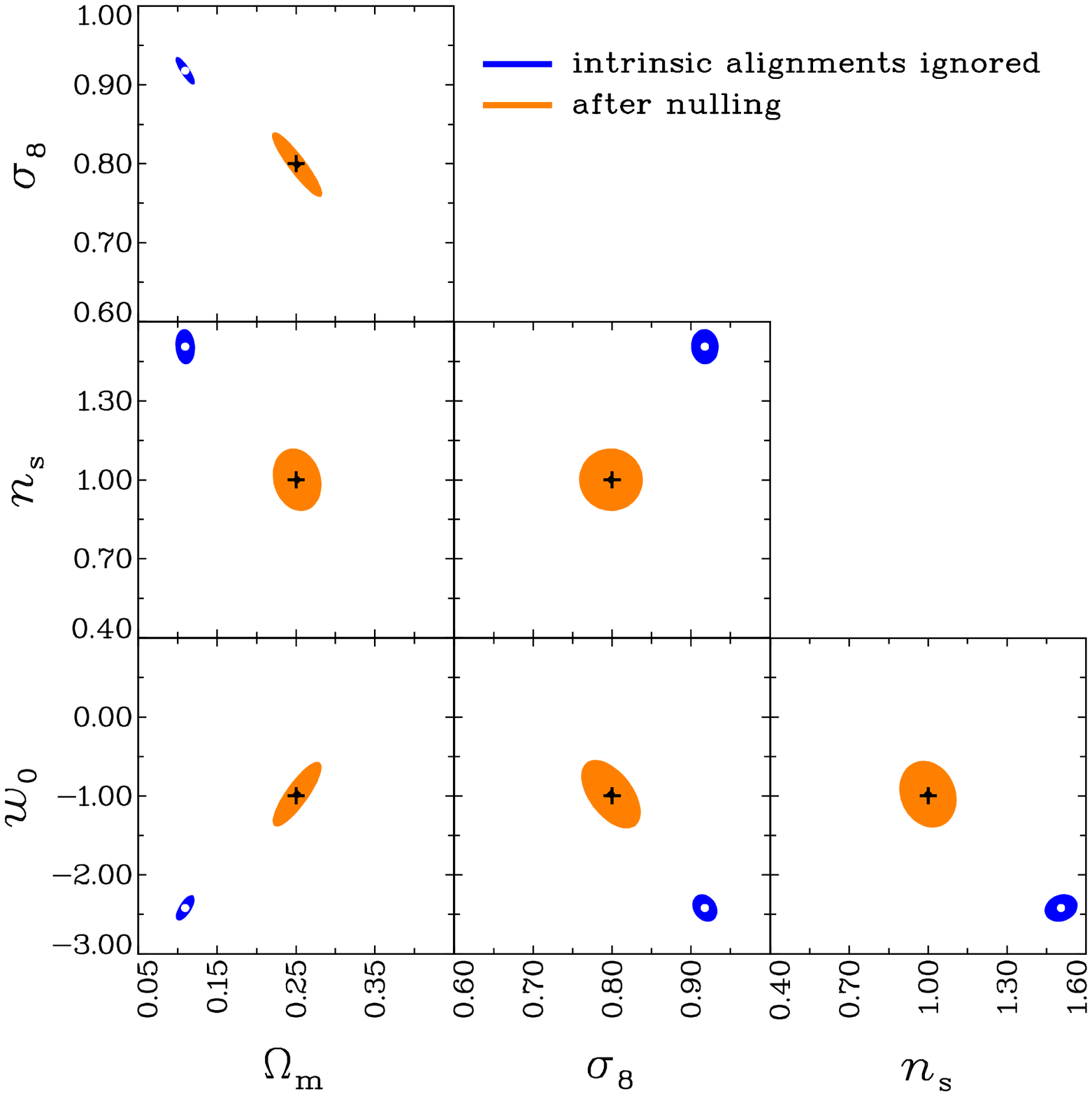}
\end{minipage}%
\begin{minipage}[c]{.28\textwidth}
\caption{Parameter constraints before and after nulling. Shown are the two-dimensional marginalised $2\sigma$-errors for the full data set (if intrinsic alignments are ignored) as blue contours and for the nulled data set as orange contours. The fiducial parameter values are marked by the crosses in each panel. The Euclid-like survey has been divided into $N_z=20$ photometric redshift bins with $\sigma_{\rm ph}=0.05$. The parameters $\bc{h,\Omega_{\rm b},w_a}$ have been marginalised over. We employed our default intrinsic alignment model.}
\label{fig:nullingfisher}
\end{minipage}
\end{figure}

The GI signal would be perfectly eliminated if the redshift bins were infinitesimally narrow. In a finite size photometric redshift bin, the redshift $z_i$, corresponding to the comoving distance $\chi_i$ where the intrinsic alignment contribution is removed, has to placed such that the GI term is still largely downweighted. Simply choosing the mean of the photometric redshift bin boundaries has proven to yield a robust and efficient nulling method, see Paper II for details. Furthermore, if the distributions $p^{(i)}(\chi)$ have finite width, the second term in (\ref{eq:limberGI}) also yields a contribution, even for $i < j$, because due to the photometric redshift scatter a galaxy from the background bin $j$ may actually be located in front of galaxies from bin $i$. This additional GI contamination is not accounted for by nulling but is limited to photometric redshift bins that are adjacent in redshift. We suppress this term by downweighting contributions of power spectra $C^{(ij)}(\ell)$ to (\ref{eq:nullingtrafo}), the stronger the smaller $j-i$, as detailed in Paper II. Note that this measure conveniently removes II correlations in the presence of photometric redshift scatter as well.

Switching back to a photometric redshift scatter $\sigma_{\rm ph}=0.05$ which should be achievable for most galaxies in future cosmic shear surveys, we predict the statistical constraints and the bias due to intrinsic alignments on a Euclid-like survey divided in to $N_z=20$ photometric redshift bins. To this end, we perform a Fisher matrix analysis on the set of cosmological parameters $\bc{\Omega_{\rm m},\sigma_8,h,n_{\rm s},\Omega_{\rm b},w_0,w_a}$, using Gaussian covariances for the power spectra that account for both cosmic variance and shape noise. The resulting $2\sigma$ confidence contours, for the case where intrinsic alignments are ignored, and after applying the nulling technique, are shown in Fig.$\,$\ref{fig:nullingfisher}. The strong intrinsic alignment contamination we have assumed in this study is reliably reduced to insignificance by nulling, including the one by the II term which is downweighted by the measures described above. The price to pay is a considerable loss in constraining power which amounts to an increase of the $1\sigma$ marginalised errors by factors of two to three on the cosmological parameters. Since nulling relies on redshift information to suppress intrinsic alignments, the dark energy equation-of-state parameters, which are primarily probed via the redshift evolution of structure growth and spacetime geometry, suffer most. It was proven in Paper II that this information loss is inevitable when summing over redshift bins as in (\ref{eq:nullingtrafo}), even if precise redshifts are available, and is a consequence of the similar redshift dependencies of the GI and GG signals, see Fig.$\,$\ref{fig:zdep}.

Since the nulling technique only makes use of the characteristic redshift dependence of the lensing and intrinsic alignment signals, its performance strongly depends on the photometric redshift accuracy. It was demonstrated in Paper II that nulling places stringent requirements on the scatter $\sigma_{\rm ph}$ and the fraction of catastrophic outliers in the relation between photometric and spectroscopic redshifts (although the nulling variant employed in this section is robust up to at least $\sigma_{\rm ph}=0.1$ and $6\,\%$ catastrophics, discarding only slightly more cosmological information if the photometric redshift quality deteriorates), as well as the uncertainty in the mean of redshift distributions. However, these requirements are met by -- or actually drive -- the goals of planned cosmic shear surveys \citep[e.g.][]{laureijs09}.

\section{Boosting intrinsic alignments}

Since in the case of intrinsic alignment boosting the weights in (\ref{eq:nullingtrafo}) are used to define the function (\ref{eq:boostingG}) over the integration range of (\ref{eq:boostingcondition}), a dense binning in redshift and hence a large number of weights ${T_{[1]}^{(i)}}_j$ is appropriate. As one does not gain information if the redshift distribution is sliced into bins which are much finer than the typical redshift scatter, the quality of redshifts drives the maximum useful number of bins and is thus again paramount for the performance of this technique. We consider three galaxy samples, one with a standard photometric redshift quality of $\sigma_{\rm ph}=0.05$ (\lq P2\rq), one with high-quality photometric redshifts $\sigma_{\rm ph}=0.03$ (\lq P1\rq), and one with spectroscopic redshifts assuming $\sigma_{\rm ph}=0$ (\lq S\rq), using all of them in combination with realistic number densities as expected for a Euclid-like survey (see Paper III for details).

\begin{figure}[t]
\begin{minipage}[c]{.6\textwidth}
\centering
\includegraphics[scale=.4,angle=270]{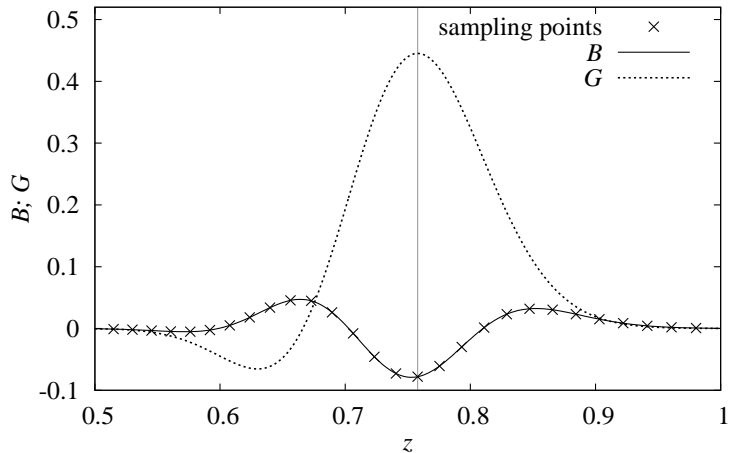}
\end{minipage}%
\begin{minipage}[c]{.4\textwidth}
\caption{Functions $G^{(i)}(\chi(z))$ and $B^{(i)}(\chi(z))$ for $z_i=0.76$, as indicated by the vertical grey line. In addition the sampling points corresponding to the median redshifts of the bins are shown for $B^{(i)}(\chi(z))$. The width of the Gaussian has been chosen to $\sigma_z=\sigma\, \dd z/\dd \chi=0.085$. Note that the normalisation of the functions is in principle arbitrary.}
\label{fig:boostingweights}
\end{minipage}
\end{figure}

We use photometric redshift bins with a separation of $0.01(1+z)$ and $0.02(1+z)$, respectively, so that one can to good approximation replace the set of weights ${T_{[1]}^{(i)}}_j$ with $j=0,\,..\,,N_z$ by a smooth weight function termed $B^{(i)}(\chi)$. In Paper III it was shown that this weight function is readily determined from the function $G^{(i)}(\chi)$ in (\ref{eq:boostingG}) via $B^{(i)}(\chi) = \chi\, \dd^2 G^{(i)}(\chi)/\dd \chi^2$. The function $G^{(i)}(\chi)$ in turn is chosen such that (\ref{eq:boostingcondition}) is fulfilled and at the same time the transformed GI signal remains strong. Among many options this can be achieved by using a parametric form
\eq{
\label{eq:defG}
\hspace*{-0.8cm} G^{(i)}(\chi) \equiv {\cal N}\; \exp \bc{-\frac{(\chi-\chi_{\rm m})^2}{\sigma^2}} \br{\chi - b}\;,
}
where ${\cal N}$, $\sigma$, $b$, and $\chi_{\rm m}$ are free parameters. These are fixed by requiring (\ref{eq:boostingcondition}), a maximum at $z_i=z(\chi_i)$ within bin $i$, and a fixed (arbitrary) normalisation. An example weight function for the sample P1 and $z_i=0.76$ is presented in Fig.$\,$\ref{fig:boostingweights}. The Gaussian in (\ref{eq:defG}) dominates the peak around $z_i$ while the term $\chi - b$ allows for the negative minimum in $G^{(i)}(\chi)$ at low redshift that is necessary to fulfil (\ref{eq:boostingcondition}). Since the corresponding weight function $B^{(i)}(\chi)$ varies considerably over the redshift range, a dense sampling by photometric redshift bins (indicated by the crosses) is required.

\begin{figure}[t]
\begin{minipage}[c]{.45\textwidth}
\centering
\includegraphics[scale=.5]{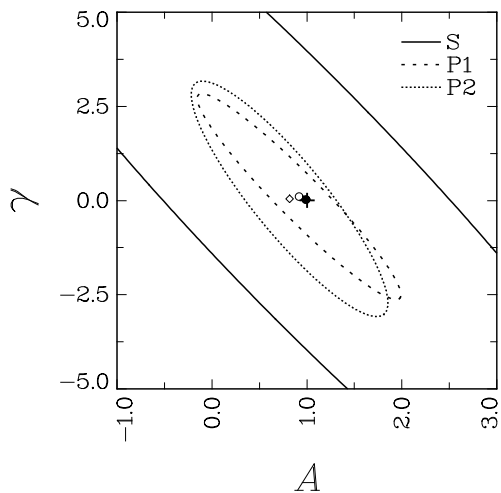}
\end{minipage}%
\begin{minipage}[c]{.55\textwidth}
\caption{Constraints on the free parameters of the GI model. Shown are the $1\,\sigma$ confidence contours for three different survey models. The solid ellipse with its centre indicated by a filled circle corresponds to the spectroscopic survey S, the dashed ellipse and open circle to survey P1, and the dotted ellipse with diamond to survey P2. Note that the centres of the contours are offset due to the bias by the residual GG signal. The cross marks the fiducial values of $A=1$ and $\gamma=0$.}
\label{fig:boostingfisher}
\end{minipage}
\end{figure}

For the same set of parameters and survey specifications we have plotted original and boosted power spectra in Fig.$\,$\ref{fig:ps}, right panel. While the power spectra $C^{(ij)}(\ell)$ have a similar level of contamination as for the case shown in the left panel of the figure, the transformed power spectrum is now clearly dominated by the GI signal, the lensing contribution being suppressed by more than an order of magnitude with respect to the GI term on all angular scales. 

To assess the signal-to-noise of these boosting-transformed power spectra, we again resort to a Fisher matrix analysis, this time constraining parameters of a toy intrinsic alignment model and treating the residual contributions by cosmic shear as the systematic that causes a bias on these parameters. We set up a two-parameter model of the form $P_{\delta {\rm I}}^{\rm model}(k,z) = A\, P_{\delta {\rm I}}(k,z)\, \bb{(1+z)/(1+z_0)}^\gamma$ with the two free parameters $A$ and $\gamma$, where $z_0$ is a pivot redshift and $P_{\delta {\rm I}}(k,z)$ is given by our default intrinsic alignment model described in Sect.$\,$\ref{sec:principle}. In Fig.$\,$\ref{fig:boostingfisher} the $1\sigma$ confidence regions for the intrinsic alignment model parameters are displayed. For all three galaxy samples that we consider the bias due to the residual GG signal is clearly subdominant to the statistical errors. Constraints by the spectroscopic sample S are very weak due to the low number density of galaxies for which spectra are available; those for the two photometric redshift samples are similar and comparable to current constraints from galaxy number density-shape cross-correlations \citep[e.g.][]{mandelbaum09,joachimi10c}. 

However, current surveys are much smaller than the one assumed for our Fisher analysis, so that the parameter constraints appear surprisingly weak. As we could demonstrate quantitatively in Paper III, the low signal-to-noise in intrinsic alignment boosting is directly related to the severe loss of information when applying the nulling technique, both governed by the difficulty of separating the GI and GG signals only by means of redshift information. Note that we did not include the II signal into the calculations of this section. Since the weight function $B^{(i)}(\chi)$ peaks close to $z_i$, see Fig.$\,$\ref{fig:boostingweights}, the II term would also be boosted with respect to the lensing signal. If one preferred to study the II and GI signals individually, the boosting transformation could be preceded by the extraction of physically close pairs of galaxies from the sample, where the efficiency of this additional step again depends critically on the quality of redshift information.

\section{Conclusions}

We demonstrated that by means of purely geometrical transformations of cosmic shear two-point statistics it is possible to both remove and extract the intrinsic alignment signal from lensing data. Both techniques robustly isolate the desired signals in a model-independent way and work for survey requirements compliant with upcoming large-area tomographic cosmic shear surveys. Since both methods make use of the characteristic dependence of the intrinsic alignment and lensing signals on redshift, the demands on (photometric) redshift quality are stringent. The performance of these approaches is fundamentally limited by the similarity of the GG and GI signals as a function of redshift, necessarily implying a considerable loss of cosmological information in the case of nulling and a low signal-to-noise in the boosting-transformed intrinsic alignment signals.

This latter finding suggests that the nulling and boosting techniques are unlikely to become the methods of choice to clean the lensing signal from intrinsic alignments or characterise intrinsic alignment signals in the galaxy sample for future cosmic shear surveys with ambitious goals on cosmological parameter constraints. However, due to their model independence and robustness, these methods can nonetheless serve as valuable cross-checks for intrinsic alignment removal methods that rely on stronger assumptions about the contamination, as well as for extrapolations of intrinsic alignment models to the redshifts and luminosities relevant for cosmic shear.

If one wishes to keep the loss of statistical power due to intrinsic alignment control in cosmic shear analyses to a minimum, one can either model the II and GI signals, incorporating a-priori information about the dependence on redshift and/or angular scales \citep[see e.g.][]{bridle07}, or make use of external data to recover parameter constraints, in particular from galaxy number density correlations and cross-correlations between galaxy number density and shapes which come for free with cosmic shear surveys \citep{bernstein08,joachimi10}.

Using simple combinations of data like in (\ref{eq:nullingtrafo}) has been identified before as an efficient, model-independent approach to eliminate potential sources of systematic errors, e.g. to remove small-scale information from the large-scale structure affected by baryonic physics \citep{huterer05} or to remove foreground contamination from cosmic microwave background multi-frequency data \citep{bennett03}. The potential benefits from these techniques for cosmic shear and related cosmological probes have not been exhausted with the present work, e.g. \citet{shi10} have applied nulling to cosmic shear three-point statistics for which the contamination by intrinsic alignments is potentially severe \citep{semboloni08}, but at present completely unknown. Extensions that allow for the inclusion of prior information on the intrinsic alignment signal, or that make use of nulling to extract the lensing magnification signal from galaxy number density-correlations are currently under investigation.

\subsection*{Acknowledgements}
We would like to thank the organisers of Astronomical Data Analysis 6, held May 2010 in Monastir, Tunisia, for a pleasant and stimulating conference. B.J. acknowledges support by the Deutsche Telekom Stiftung and the Bonn-Cologne Graduate School of Physics and Astronomy.

\bibliographystyle{aa}

\end{document}